\documentclass[prd,aps,showpacs,superscriptaddress,nofootinbib,amsmath,amssymb]{revtex4}

\usepackage{graphicx}% Include figure files
\usepackage{dcolumn}% Align table columns on decimal point
\usepackage{bm}% bold math
\usepackage{lipsum}
\begin{document}
\title{Nuclear medium effects in Drell-Yan process}
\author{H. Haider}
\author{M. Sajjad Athar}
\email{sajathar@gmail.com}
\author{S. K. Singh}
\affiliation{Department of Physics, Aligarh Muslim University, Aligarh - 202 002, India}
\author{I. \surname{Ruiz Simo}}
\affiliation{Departamento de F\'{\i}sica At\'omica, Molecular y Nuclear,
and Instituto de F\'{\i}sica Te\'orica y Computacional Carlos I,
Universidad de Granada, Granada 18071, Spain}
\begin{abstract}
We study the nuclear medium effects in Drell-Yan process using quark parton distribution functions calculated in a microscopic nuclear model which takes into account the 
effects of Fermi
 motion, nuclear binding and nucleon correlations through a relativistic nucleon spectral function. The contributions of $\pi$ and $\rho$ mesons as well as 
 shadowing effects are also included.
 The beam energy loss is calculated using a phenomenological approach. The present theoretical results are compared with the experimental results of E772 and E886 experiments. These results are applicable to the forthcoming experimental analysis of E906 Sea
Quest experiment at Fermi Lab.
\end{abstract}
\maketitle
\section{Introduction} 
\begin{figure}
\includegraphics[scale=0.8]{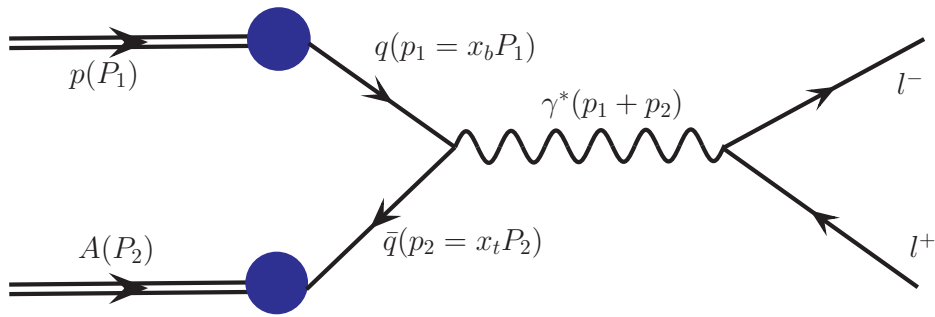}
\caption{Drell-Yan process: Here p stands for a proton and A for a proton or a nucleus. In the brackets four momenta of the particles are mentioned.}
\label{fig0}  
\end{figure}

Drell-Yan(DY) production of lepton pairs~\cite{drellyan} from nucleons and nuclear targets is an important tool to study the quark structure 
of nucleons and its modification in the nuclear
medium. In particular, the proton induced DY production of muon pairs on nucleons and nuclei provides a direct probe
to investigate the quark parton
distribution functions(PDFs). In a DY process(shown in Fig.\ref{fig0}), a quark of beam(target) hadron gets annihilated from the antiquark of target(beam) hadron and gives rise 
to a photon which in turn gives lepton pairs of opposite charge. The basic process is 
$q^{b(t)}+\bar q^{t(b)}\rightarrow l^++l^-$, where b and t indicate the beam proton and the target nucleon/hadron. A quark(antiquark) 
in the beam carrying a longitudinal momentum fraction $x_b$  interacts with an antiquark(quark) in the target carrying longitudinal momentum fraction $x_t$
of the target
momentum per nucleon to produce a virtual photon. 
The cross section per target nucleon $\frac{d^{2}\sigma}{dx_bdx_t}$ in the leading order is given by~\cite{Amanda}:
\begin{eqnarray}\label{DY1}
\frac{d^2\sigma}{dx_bdx_t}&=&\frac{4\pi\alpha^2}{9 Q^2}\sum_fe_f^{2}\left\{q_{f}^{b}(x_b,Q^2)\bar q_{f}^{t}(x_t,Q^2) + \bar q_{f}^{b}(x_b,Q^2)q_{f}^{t}(x_t,Q^2)\right\}
\end{eqnarray}
where $\alpha$ is the fine structure constant, $e_f$ is the charge of quark/antiquark of flavor f, $Q^2$ is the photon virtuality and $q_{f}^{b(t)}(x)$
and $\bar q_{f}^{b(t)}(x)$ are the beam(target) quark/antiquark PDFs of flavour f.

This process is directly sensitive to the antiquark parton distribution functions $\bar q(x)$ in target nuclei which has also 
 been studied by DIS experiments through the observation of EMC effect. Quantitatively the EMC effect describes the nuclear modification of nucleon structure
 function $F_2(x_t)$ for 
 the bound nucleon defined as $F_2(x_t)=x_t\sum_fe_f^2[q_f(x_t)~+~\bar q_f(x_t)]$ and gives information about the modification of the sum of quark and antiquark
 PDFs~\cite{kenyon,Geesaman:1995yd} which is
  dominated by the valence quarks in the high $x_t$ region ($x_t > 0.3$). In the low $x_t$ region ($x_t \le 0.3$), where sea quarks are expected to give 
  dominant contribution, the study of $F_2(x_t)$ gives
   information about sea quark and antiquark PDFs. Thus, nuclear modifications are phenomenologically incorporated in $q(x_t)$ and $\bar q(x_t)$ using the
   experimental data on $F_2(x_t)$ and are used to analyze the DY yields from nuclear targets. Some authors succeed in giving a 
   satisfactory description of DIS and DY data on nuclear targets using same set of nuclear $q(x)$ and $\bar q(x)$~\cite{eskola09}, while some others find
   it difficult to provide a consistent description
   of DIS and DY data using the same set of nuclear PDFs~\cite{scheinbein2008}. On the other hand, there are many theoretical attempts to describe the nuclear 
   modifications of quark and antiquark
    PDFs to explain DIS which have also been used to understand the DY process on nuclear targets
    ~\cite{miller}-\cite{Marco:1997xb}. The known nuclear modifications discussed in literature in the case of DIS are (a) modification of nucleon structure 
    inside the nuclear medium, (b) a significantly enhanced 
    contribution of subnucleonic degrees of freedom like pions or quark clusters in nuclei and (c) nuclear shadowing. 
    \begin{figure}
\includegraphics[scale=0.5]{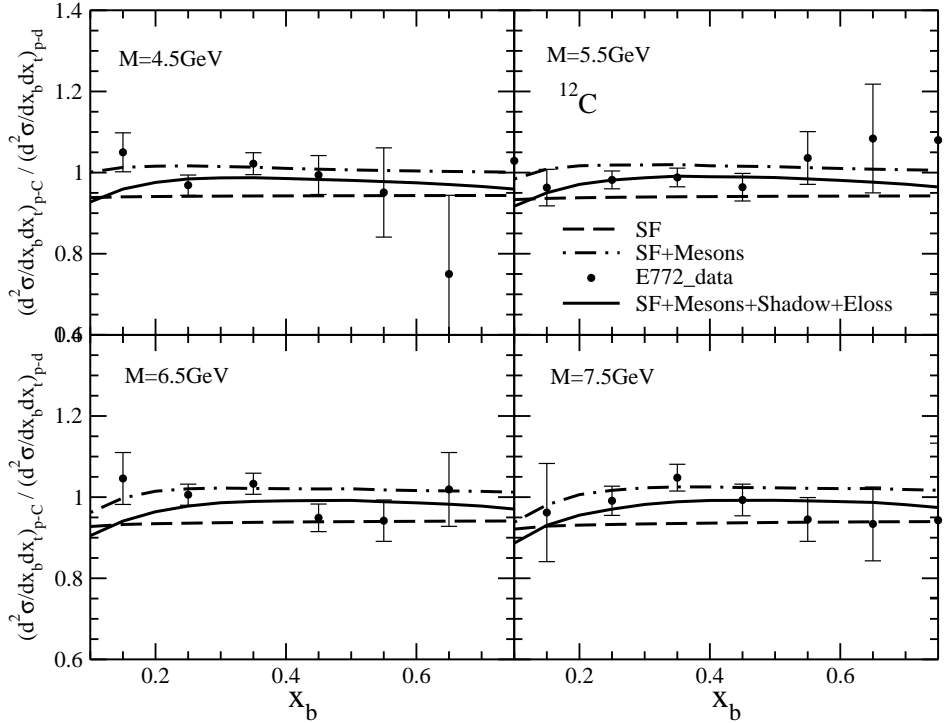}
\caption{$\frac{\left(\frac{d\sigma}{dx_b~dx_t}\right)_{p - ^{^{12}}C}}{\left(\frac{d\sigma}{dx_b~dx_t}\right)_{p - ^{^2}D}}$ vs $x_b$ at
 E=800GeV($\sqrt{s_N}$=38.8GeV). 
 Spectral function: dashed line, including mesonic contribution: dashed-dotted line and the results obtained using the full model i.e. 
 spectral function+meson cloud contributions+shadowing effects+energy loss: solid line. The results in the different columns are obtained at different values of 
 M(=$\sqrt {Q^2}$). Experimental points are data of E772 experiment~\cite{alde,dyhepdata}.}
\label{fig1}  
\end{figure}
      However, in the case of DY processes there is an additional nuclear effect due to initial state interaction of beam partons with the target partons which may be 
      present before the hard 
      collisions  of these partons giving rise to lepton pairs. As the initial beam traverses the nuclear medium it loses energy due to interaction of beam partons 
      with nuclear constituents of the target. This can be 
      visualized in terms of the
     interaction of hadrons or its constituents with the constituents of the target nucleus through various inelastic processes leading to energy loss of the 
     interacting beam partons. This has been 
     studied phenomenologically using available parameterization of nuclear PDFs or theoretically in models based on QCD or Glauber approaches taking into account the 
     effect of shadowing which also plays an important role in the low $x_t$ region,  however, any consensus in the understanding of physics behind the beam energy loss
     has been lacking. In this scenario most of the calculations incorporate a phenomenological description 
     of beam energy loss to explain the experimental data on DY yields~\cite{vasilev,Duan2005,Johnson2002,Johnson:2000ph,Arleo:2002ph,Garvey2003,Brodsky1993}.
          In this region of $x_t$ the nuclear modification of sea quark PDF and mesonic contributions also become important. 
%            It is, however, known that mesonic 
%            contributions enhance DY yields
%            (and $F_2(x)$ in DIS) 
%       while the shadowing and parton energy loss effects suppress them.
Thus in this process, main nuclear effects 
      are due to nuclear structure, mesonic contributions and shadowing (as in the case of DIS) with additional effect of parton energy loss in the beam parton energy due to the presence of nuclear targets. 
  
 \begin{figure}
\includegraphics[scale=0.5]{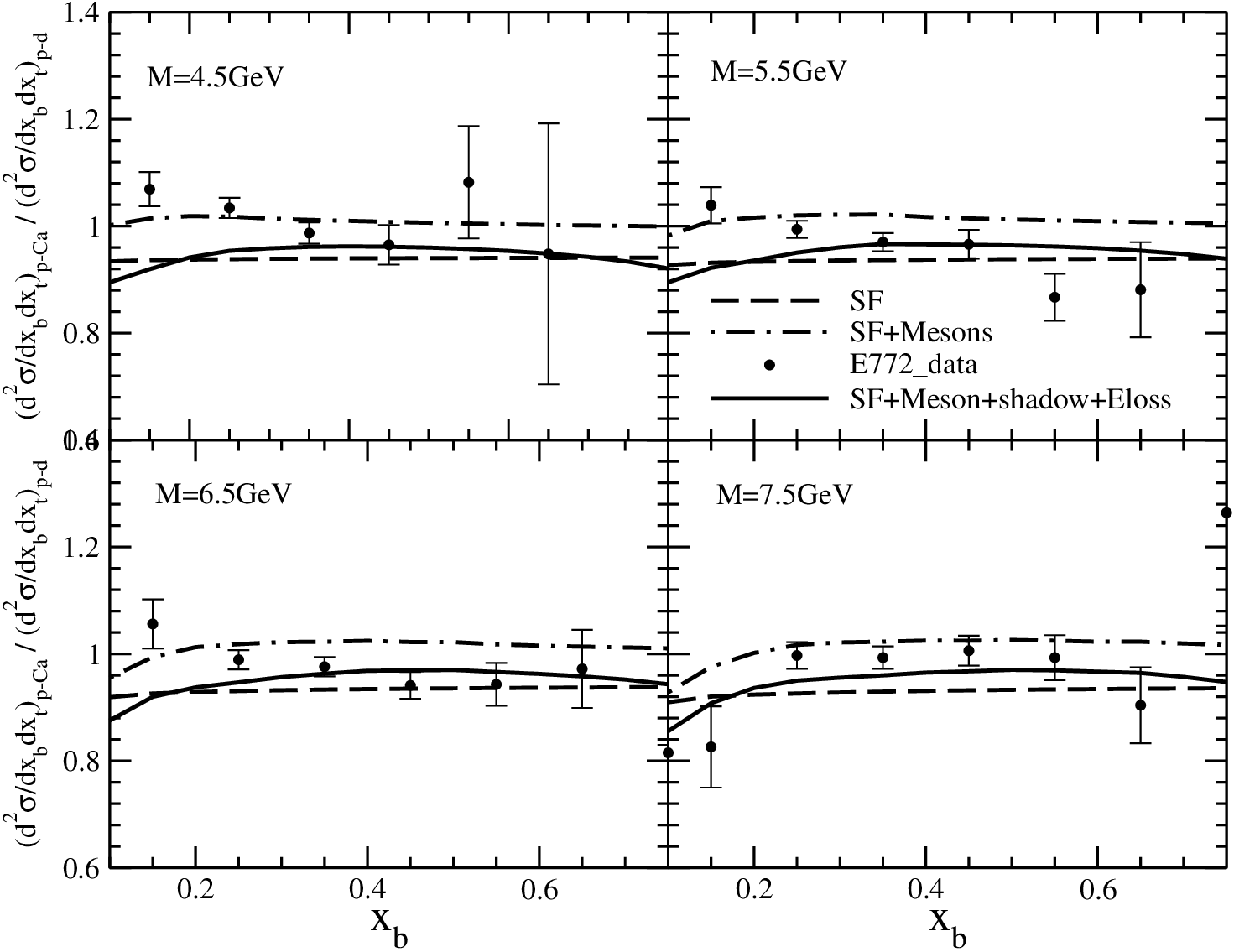}
\caption{$\frac{\left(\frac{d\sigma}{dx_b~dx_t}\right)_{p - ^{^{40}}Ca}}{\left(\frac{d\sigma}{dx_b~dx_t}\right)_{p - ^{^2}D}}$
 vs $x_b$ at E=800GeV($\sqrt{s_N}$=38.8GeV). Lines and points have 
the same meaning as in Fig.\ref{fig1}}
\label{fig2}  
\end{figure}
       In this paper, we present the results of nuclear medium effects on DY production
        of lepton pairs calculated in a microscopic nuclear model which has been successfully used to describe the DIS of charged leptons and $\nu$/$\bar\nu$ 
        from various nuclei
~\cite{marco1996,sajjadnpa,prc84,prc85,Haider:2015vea,Haider:2016zrk}. The model uses a relativistic nucleon spectral function to describe target nucleon momentum distribution incorporating Fermi motion, binding energy 
effects and nucleon correlations in a field theoretical model. The model has also been used to include the mesonic contributions from $\pi$ and $\rho$ mesons. 
The beam energy loss has been calculated
using some phenomenological models discussed in the literature~\cite{Duan2005}-\cite{vasilev}. The results have been presented for the 
kinematic region of experiments E772~\cite{alde} and E866~\cite{e866,vasilev}  or proton induced DY processes in nuclear
targets like $^{9}Be$, $^{12}C$, $^{40}Ca$, $^{56}Fe$ and $^{184}W$ in the region of $x_t > 0.1$. 
The numerical results extended up to $x_t=0.45$, should be useful in analyzing the forthcoming experimental results from the SeaQuest E906 experiment 
being done at Fermi Lab~\cite{Seaquest}.

     In section-\ref{sec:NE}, we present the formalism in brief;  in section-\ref{sec:RD}, the results are presented and discussed;  and 
finally in section-\ref{sec:CC}, we summarize the results and conclude our findings. 

\begin{figure}
\includegraphics[scale=0.5]{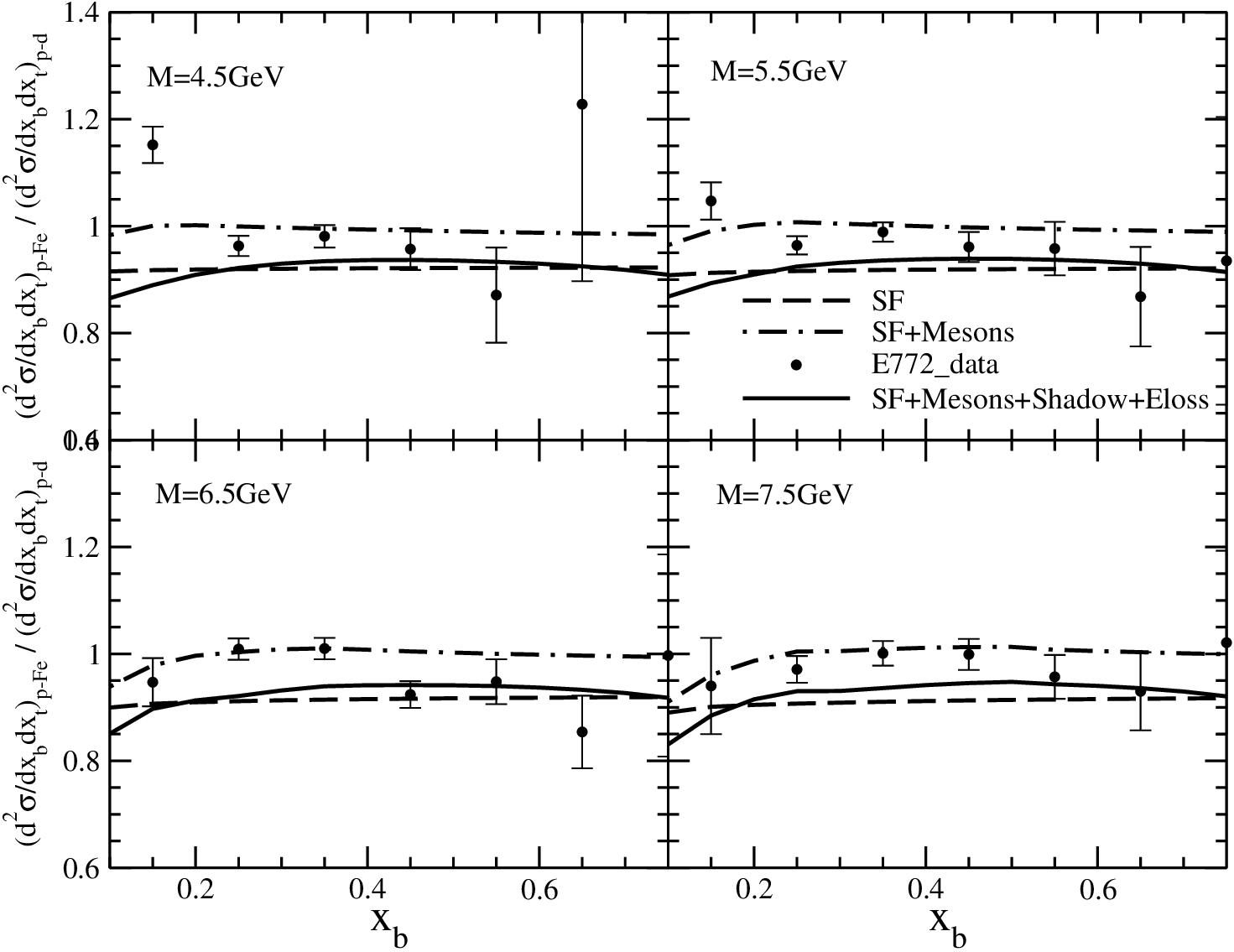}
\caption{$\frac{\left(\frac{d\sigma}{dx_b~dx_t}\right)_{p - ^{^{56}}Fe}}{\left(\frac{d\sigma}{dx_b~dx_t}\right)_{p - ^{^2}D}}$ 
 vs $x_b$ at E=800GeV($\sqrt{s_N}$=38.8GeV). Lines and points have 
the same meaning as in Fig.\ref{fig1}}
\label{fig3}  
\end{figure}

\section{Nuclear effects}\label{sec:NE}
When DY process takes place in nuclei, nuclear effects appear which are generally believed to be due to \\
(a) nuclear structure effects arising from Fermi motion, binding energy and nucleon correlations,\\
(b) additional contribution due to subnucleonic degrees of freedom like mesons and/or quark clusters in the nuclei, \\
(c) shadowing effect,
and\\
(d) energy loss of the beam proton as it traverses the nuclear medium before producing lepton pairs.

In the case of proton induced DY processes in nuclei, the target nucleon has a Fermi momentum described by a momentum distribution. 
The target Bjorken variable $x_t$ is defined for a free nucleon 
 as $x_t=\frac{2q.p_1}{(p_1+p_2)^2}$, where q is the four momentum of $\mu^+\mu^-$ pair, $p_{1\mu}$ and $p_{2\mu}$ are respectively the beam and target four momenta 
in the nuclear medium. Moreover, the projectile Bjorken variable $x_b$ expressed covariantly as $x_b=\frac{2 q.p_2}{(p_1+p_2)^2}$
also changes due to the energy loss of the beam particle caused by the initial state interactions with the 
nuclear constituents as it travels through the nuclear medium before producing lepton pairs.
These nuclear modifications are incorporated while evaluating Eq.(\ref{DY1}). Furthermore, there are 
additional contributions from the pion and rho mesons which are also taken into account.

In the following, we briefly outline the model and refer to earlier 
work~\cite{Marco:1997xb,marco1996,sajjadnpa} for details.              
%\cite{Marco:1997xb}\cite{marco1996}-\cite{prc87}.

\subsection{Nuclear Structure}
In a nucleus, scattering is assumed to take place from partons inside the individual nucleons which are bound and moving with a momentum $\vec p$
within a limit given by the Fermi momentum. The target Bjorken variable 
$x_t$ becomes Fermi momentum dependent and PDF for quarks and antiquarks in the nucleus i.e. $q_f^t(x_t)$ and $\bar q_f^t(x_t)$ are calculated as a convolution of the PDFs 
in bound nucleon and a momentum distribution function of the nucleon inside the nucleus. The parameters of the momentum distribution are adjusted to correctly incorporate nuclear properties
 like binding energy, Fermi motion and the nucleon correlation effects in the nuclear medium. We use the Lehmann representation of the relativistic Dirac propagator for an interacting Fermi sea in nuclear 
 matter to derive such a momentum distribution and  Local Density Approximation to translate these results for a finite nucleus~\cite{Marco:1997xb,sajjadnpa,prc84, prc85,Haider:2015vea}. The free relativistic propagator for a nucleon of mass $M_N$ is written in terms of 
 positive and negative energy components as
\begin{equation}  \label{prop2}
G^{0}(p_{0},{\bf p}) =\frac{M_N}{E({\bf p})}\left\{\frac{\sum_{r}u_{r}({\bf p})\bar u_{r}({\bf p})}{p^{0}-E({\bf p})+i\epsilon}+\frac{\sum_{r}v_{r}(-{\bf p})\bar v_{r}(-{\bf p})}{p^{0}+E({\bf p})-i\epsilon}\right\}
\end{equation} 
For a noninteracting Fermi sea where only positive energy solutions are considered the relevant propagator is rewritten in terms of occupation number $n({\bf p})=1$
 for p$\le p_{F}$ while $n({\bf p})$=0 for p$> p_{F}$:
\begin{eqnarray}  \label{prop4}
G^{0}(p_{0},{\bf p})&=&\frac{M_N}{E({\bf p})}\left\{\sum_{r}u_{r}({\bf p})\bar u_{r}({\bf p})\left[\frac{1-n({\bf p})}{p^{0}-E({\bf p})+i\epsilon}+\frac{n({\bf p})}{p^{0}-E({\bf p})-i\epsilon}\right]\right\}
\end{eqnarray}
 The nucleon propagator in an interacting Fermi sea is then calculated by making a perturbative expansion of $G(p_{0},{\bf p})$ in terms of free nucleon propagator
  $G^{0}(p_{0},{\bf p})$ given 
 in Eq.~(\ref{prop2}) by retaining the positive energy contributions only (the negative energy components are suppressed).

\begin{figure}
\includegraphics[scale=0.5]{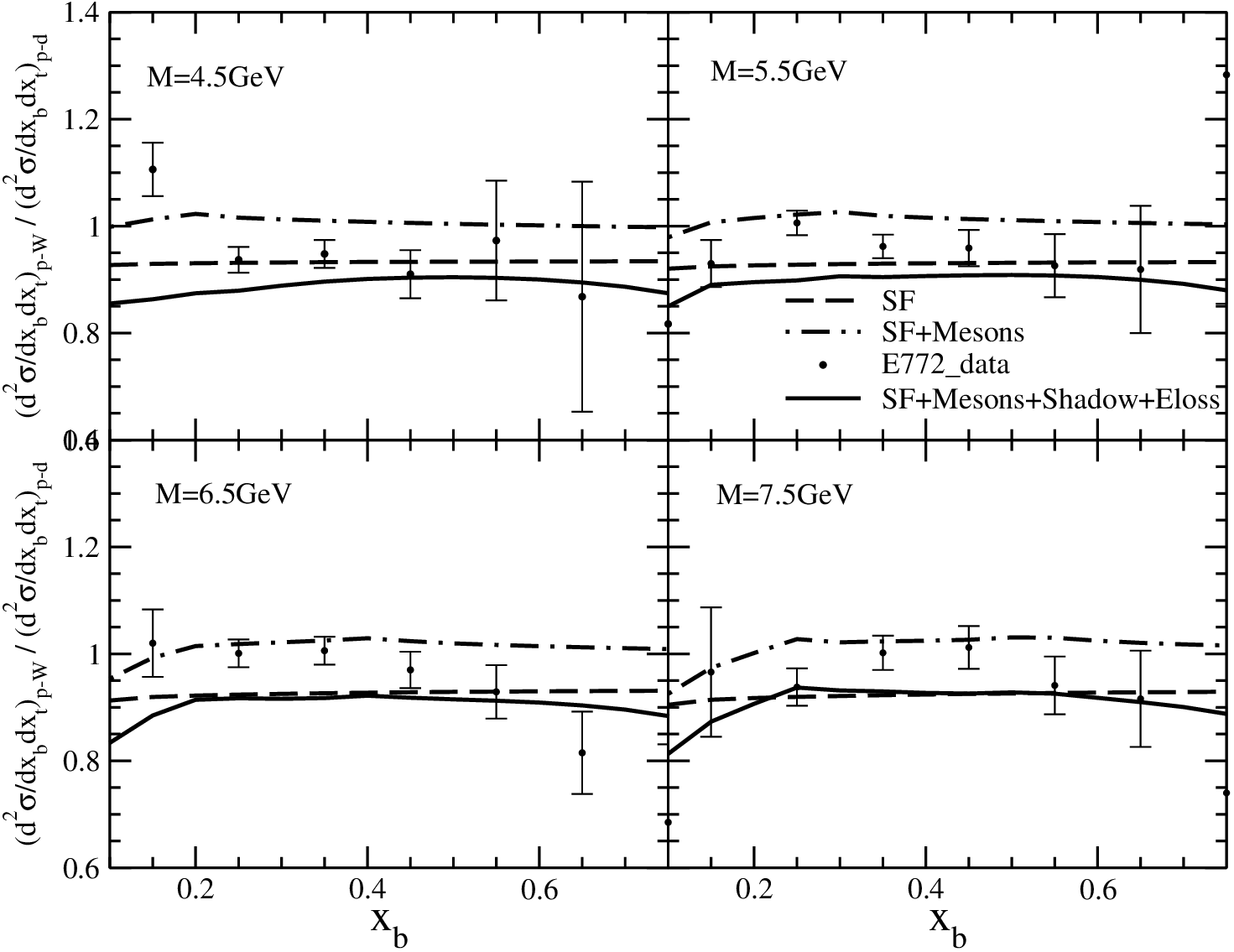}
\caption{$\frac{\left(\frac{d\sigma}{dx_b~dx_t}\right)_{p - ^{^{184}}W}}{\left(\frac{d\sigma}{dx_b~dx_t}\right)_{p - ^{^2}D}}$ vs $x_b$ at E=800GeV($\sqrt{s_N}$=38.8GeV).  Lines and points have 
the same meaning as in Fig.\ref{fig1}}
\label{fig4}  
\end{figure}

This perturbative expansion is then summed in ladder approximation to give dressed nucleon propagator $G(p_{0},{\bf p})$ ~\cite{marco1996,FernandezdeCordoba:1991wf}
\begin{eqnarray}
G(p_{0},{\bf p})&=&\frac{M_N}{E({\bf p})}\sum_{r}u_{r}({\bf p})\bar u_{r}({\bf p})\frac{1}{p^{0}-E({\bf p})}+\frac{M_N}{E({\bf p})}
\sum_{r}\frac{u_{r}({\bf p})\bar u_{r}({\bf p})}{p^{0}-E({\bf p})}\sum(p^{0},{\bf p})\frac{M_N}{E({\bf p})}\sum_{s}\frac{u_{s}({\bf p})\bar u_{s}({\bf p})}{p^{0}-E({\bf p})}+.....\nonumber \\
&&=\frac{M_N}{E({\bf p})}\frac{\sum_{r} u_{r}({\bf p})\bar u_{r}({\bf p})}{p^{0}-E({\bf p})-\sum(p^{0},{\bf p})\frac{M_N}{E({\bf p})}},
\end{eqnarray}
where $\sum(p^{0},{\bf p})$ is the nucleon self energy.

This allows us to write the relativistic nucleon propagator in a nuclear medium in terms of the Spectral functions of hole and 
particle as~\cite{FernandezdeCordoba:1991wf}
\begin{eqnarray}\label{Gp}
G (p^0, {\bf p})=\frac{M_N}{E({\bf p})} 
\sum_r u_r ({\bf p}) \bar{u}_r({\bf p})
\left[\int^{\mu}_{- \infty} d \, \omega 
\frac{S_h (\omega, \bf{p})}{p^0 - \omega - i \eta}
+ \int^{\infty}_{\mu} d \, \omega 
\frac{S_p (\omega, \bf{p})}{p^0 - \omega + i \eta}\right]\,
\end{eqnarray}
where $S_h (\omega, \bf{p})$ and $S_p (\omega, \bf{p})$ being the hole
and particle spectral functions respectively, which are derived in Ref.~\cite{FernandezdeCordoba:1991wf}, and $\mu$ is the chemical potential.
We use:
\begin{equation}\label{sh}
S_h (p^0, {\bf p})=\frac{1}{\pi}\frac{\frac{M_N}{E({\bf p})}Im\Sigma(p^0,{\bf p})}{(p^0-E({\bf p})-\frac{M_N}{E({\bf p})}Re\Sigma(p^0,{\bf p}))^2
+ (\frac{M_N}{E({\bf p})}Im\Sigma(p^0,{\bf p}))^2}
\end{equation}
for $p^0 \le \mu$
\begin{equation}\label{sh1}
S_p (p^0, {\bf p})=-\frac{1}{\pi}\frac{\frac{M_N}{E({\bf p})}Im\Sigma(p^0,{\bf p})}{(p^0-E({\bf p})-\frac{M_N}{E({\bf p})}Re\Sigma(p^0,{\bf p}))^2
+ (\frac{M_N}{E({\bf p})}Im\Sigma(p^0,{\bf p}))^2}
\end{equation}
for $p^0 > \mu$.

The normalization of this spectral function is obtained by imposing the baryon number conservation following the method of Frankfurt and Strikman~\cite{Frankfurt}.
 In the present paper, we use local density approximation (LDA) where we do not have a box of constant density, and the reaction takes 
place at a point ${\bf r}$, lying inside a volume element $d^3r$ with local density $\rho_{p}({\bf r})$ and $\rho_{n}({\bf r})$ 
corresponding to the proton and neutron densities at the point ${\bf r}$. This leads to the spectral functions for the protons and 
neutrons to be the function of local Fermi momentum given by
 \begin{eqnarray} \label{Fermi1}
k_{F_p}({\bf r})= \left[ 3\pi^{2} \rho_{p}({\bf r})\right]^{1/3},~~k_{F_n}({\bf r})= \left[ 3\pi^{2} \rho_{n}({\bf r})\right]^{1/3}
\end{eqnarray}
and therefore the normalization condition may be imposed as
\begin{eqnarray} \label{norm2}
2\int\frac{d^{3}p}{(2\pi)^{3}}\int_{-\infty}^{\mu_{p(n)}}S_{h}^{p(n)}(\omega,{\bf p},k_{F_{p,n}}({\bf r})) d\omega= \rho_{p,n}({\bf r}),
\end{eqnarray}
where the factor of two is to take into account spin degrees of freedom of proton and neutron, and $\mu_{p}$ and $\mu_{n}$ are the chemical potentials for proton and neutron respectively. 

 This further leads to the normalization condition given by
\begin{equation}\label{norm4}
2 \int d^3 r \;  \int \frac{d^3 p}{(2 \pi)^3} 
\int^{\mu_{p(n)}}_{- \infty} \; S_h^{p(n)} (\omega, {\bf p}, \rho_{p(n)}({\bf r})) 
\; d \omega = Z(N)\,,
\end{equation}

The average kinetic and total nucleon energy in a nucleus with the same number of protons and neutrons are given by: 
\begin{eqnarray}
<T>= \frac{4}{A} \int d^3 r \;  \int \frac{d^3 p}{(2 \pi)^3} (E({\bf p})-M_N) 
\int^{\mu}_{- \infty} \; S_h (p^0, {\bf p}, \rho(r)) 
\; d p^0\,,
\end{eqnarray}
\begin{eqnarray}
<E>= \frac{4}{A} \int d^3 r \;  \int \frac{d^3 p}{(2 \pi)^3}  
\int^{\mu}_{- \infty} \; S_h (p^0, {\bf p}, \rho(r)) 
\; p^0 d p^0\,,
\end{eqnarray}
where $\rho(r)$ is the baryon density for the nucleus which is normalized to A and is taken from the electron nucleus scattering experiments.
The binding energy per nucleon is given by~\cite{marco1996}:
\begin{equation}\label{be}
|E_A|=-\frac{1}{2}(<E-M_N>+\frac{A-2}{A-1}<T>)
\end{equation}
The binding energy per nucleon for each nucleus is correctly reproduced to match with the experimentally observed values. 
 Once the spectral function is normalized to the number of nucleons and we obtain the correct binding energy, there is no free parameter that is left in
 our model.% to take into account nuclear medium like Fermi motion, binding energy and nucleon correlations.

 In the case of nucleus, the nuclear hadronic tensor $W^{\mu \nu}_A$ for an isospin symmetric nucleus is derived to be~\cite{marco1996,sajjadnpa}:
\begin{eqnarray}\label{w2Anuclei}
W^{\mu \nu}_A&=& 2 \sum_{i=p,n} \int \, d^3 r \, \int \frac{d^3 p}{(2 \pi)^3} \, 
\frac{M_N}{E (\vec{p})} \, \int^{\mu}_{- \infty} d p^0 S_h (p^0, {\bf p},\rho_{i}) W^{\mu \nu}_i (p, q) 
\end{eqnarray}
Using this, the electromagnetic structure function $F_{2A}(x,Q^2)$ for a non-symmetric (N$\ne$Z) nucleus in DIS is obtained as~\cite{marco1996}, 
\begin{eqnarray}\label{f2Anuclei}
F_{2A}(x,Q^2)&=&2\sum_{i=p,n}\int d^3r\int\frac{d^3p}{(2\pi)^3}\frac{M_N}{E(\mathbf{p})}\int^{\mu_i}_{-\infty}dp^0\; 
S^{i}_{h}(p^0,\mathbf{p},\rho_i(r)) 
\sum_f e_f^2 x_t^\prime [q_f^i(x_t^\prime(p^0, {\vec p}))+\bar q_f^i(x_t^\prime(p^0, {\vec p}))] \nonumber\\
\end{eqnarray}
  For the numerical calculations, we have used CTEQ6.6~\cite{cteq} nucleon parton distribution functions(PDFs) for quark($q_f^i$) and antiquark($\bar q_f^i$) of flavor f. 
 
Following the same procedure as taken for the evaluation of nuclear structure function, we incorporate the nuclear medium effects 
like Fermi motion, binding energy and nucleon correlations in the evaluation of bound quarks in nucleons of a nucleus. 
$q_{f}^t(x_t)$ and $\bar q_{f}^t(x_t,Q^2)$ are expressed in terms of spectral function as~\cite{Marco:1997xb}:
\begin{eqnarray}\label{qbarA}
q_{f}^t(x_t,Q^2)&=&2\sum_{i=p,n}\int d^3r\int\frac{d^3p}{(2\pi)^3}\frac{M_N}{E(\mathbf{p})} \int^{\mu_i}_{-\infty}dp^0\; 
S_{h}^i(p^0,\mathbf{p},\rho_i(r)) {q}_f^i(x_t^\prime(p^0, {\vec p}),Q^2)\nonumber\\
\bar q_{f}^t(x_t,Q^2)&=&2\sum_{i=p,n}\int d^3r\int\frac{d^3p}{(2\pi)^3}\frac{M_N}{E(\mathbf{p})} \int^{\mu_i}_{-\infty}dp^0\;
S_{h}^i(p^0,\mathbf{p},\rho_i(r)) {\bar q}_f^i(x_t^\prime(p^0, {\vec p}),Q^2),
\end{eqnarray}
 where ${q}_f^i(\bar q_{f}^i(x_t,Q^2))$ is the quark(antiquark) PDFs for flavor f inside a nucleon of kind i and the factor of 2 is because of quark(antiquark) spin degrees of freedom. ${x_t}^\prime = \frac{M_N}{p^0 - p_z} x_t$ which is obtained 
 from the covariant expression of $x_t^\prime=\frac{q\cdot p_1}{s_N}$ with $q\| z$ direction.
 
\begin{figure}
\includegraphics[scale=0.5]{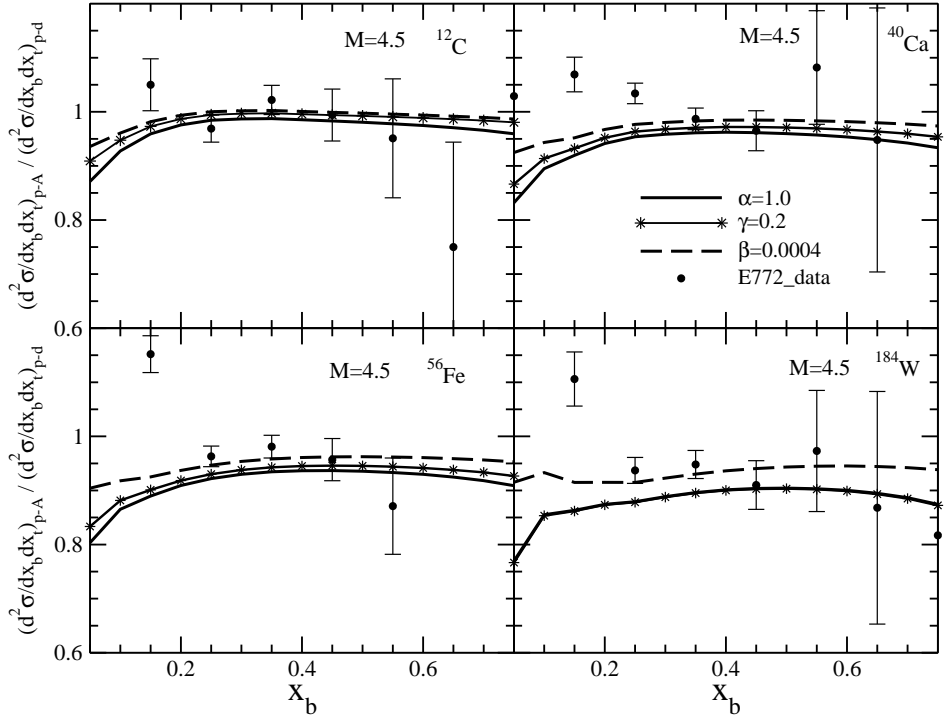}
\caption{$\frac{\left(\frac{d\sigma}{dx_b~dx_t}\right)_{p - A}}{\left(\frac{d\sigma}{dx_b~dx_t}\right)_{p - ^{^2}D}}$ vs $x_b$ at 
E=800GeV($\sqrt{s_N}$=38.8GeV). The results are 
obtained using the full model at M=4.5GeV. 'A' stands for several nuclei like  $^{12}C$, $^{40}Ca$, $^{56}Fe$ and $^{184}W$. 
 These results are obtained using different models 
for the energy loss viz. $\alpha=1$ in Eq.(\ref{alpha}) shown by the solid line, $\gamma=0.2$ in Eq.(\ref{aalpha}) shown by the solid line with stars and 
$\beta=0.0004$ in Eq.(\ref{beta}) shown by the dashed line.
 Experimental points are data of E772 experiment~\cite{alde,dyhepdata}.}
\label{fig66}  
\end{figure}

\subsection{Mesonic contributions}
As the nucleons are strongly interacting particles and inside the nucleus continuous exchange of virtual mesons take place, therefore, we have also taken into account
the probability of interaction of virtual photons with the meson clouds. In the present work, we have considered $\pi$ and $\rho$ mesons.
For this the imaginary part of the meson propagators are introduced instead of spectral function which were derived 
from the imaginary part of the
 nucleon propagator. Therefore, in the case of pion, we replace in Eq.(\ref{f2Anuclei})~\cite{sajjadnpa}:
 \[\frac{M_N}{E(\mathbf{p})}\int^{\mu}_{-\infty}d\omega\; S_{h}(\omega,\mathbf{p})\; \delta(p^0-\omega)\rightarrow~-\frac{1}{\pi}~\theta(p_0)\; ImD(p)\]
 where $D(p)$ is the pion propagator in the nuclear medium given by 
 \begin{equation}\label{dpi}
D (p) = [ {p^0}^2 - \vec{p}\,^{2} - m^2_{\pi} - \Pi_{\pi} (p^0, {\bf p}) ]^{- 1}\,,
\end{equation}
with
\begin{equation}\label{pionSelfenergy}
\Pi_\pi=\frac{f^2/m_\pi^2 F^2(p)\vec{p}\,^{2}\Pi^*}{1-f^2/m_\pi^2 V'_L\Pi^*}\,.
\end{equation}
Here, $F(p)=(\Lambda_{\pi}^2-m_\pi^2)/(\Lambda_{\pi}^2+\vec{p}\,^{2})$ is the $\pi NN$ form factor, $\Lambda_{\pi}$=1GeV, $f=1.01$, $V'_L$ is
the longitudinal part of the spin-isospin interaction and $\Pi^*$ is the irreducible pion self energy that contains the contribution of particle - hole and delta - hole excitations.

Following a similar procedure, as done in the case of nucleon, the contribution of the pions to hadronic tensor in the nuclear medium may be written as \cite{marco1996}
\begin{equation}\label{W2pion}
W^{\mu \nu}_{A, \pi} = 3 \int d^3 r \; \int \frac{d^4 p}{(2 \pi)^4} \;
\theta (p^0) (- 2) \; Im D (p) \; 2 m_\pi W^{\mu \nu}_{\pi} (p, q)
\end{equation}
However, Eq.(\ref{W2pion}) also contains the contribution of the pionic contents of the nucleon, which are already contained in the sea contribution of nucleon through Eq.(\ref{qbarA}), therefore, the pionic 
contribution of the nucleon is to be subtracted from Eq.(\ref{W2pion}), in order to calculate the contribution from the excess pions in the nuclear medium. This is obtained by replacing $I m D (p)$ by 
$\delta I m D (p)$~\cite{marco1996} as
\begin{equation}
Im D (p) \; \rightarrow \; \delta I m D (p) \equiv I m D (p) - \rho \;
\frac{\partial Im D (p)}{\partial \rho} \left|_{\rho = 0} \right.
\end{equation}

Using Eq.(\ref{W2pion}), pion structure function $F_{2, \pi}^A(x)$ in a nucleus is derived as
\begin{equation}  \label{F2pion}
F_{2, \pi}^A (x) = - 6 \int  d^3 r  \int  \frac{d^4 p}{(2 \pi)^4} \; 
\theta (p^0) \; \delta I m D (p) \; 
\; \frac{x}{x_\pi} \; 2 M_N \; \sum_f e_f^2 x_\pi [q^f_{\pi}(x_\pi(p^0, {\vec p}))+\bar q^f_{\pi}(x_\pi(p^0, {\vec p}))]     \; \theta (x_\pi - x) \; 
\theta (1 - x_\pi), 
\end{equation}
where $\frac{x}{x_{\pi}} = \frac{- p^0 + p^z}{M_N}$.

This in turn leads to the expression for the pion quark PDF in the nuclear medium. For example, $q_{f,\pi}^t(x_t,Q^2)$ is derived as~\cite{Marco:1997xb}:
\begin{eqnarray}  \label{F2piqbar}
q_{f,\pi}^t(x_t,Q^2) = - 6 \int  d^3 r  \int  \frac{d^4 p}{(2 \pi)^4} \theta (p^0) \delta I m D (p)  2 M_N  q_{f,\pi}(x_\pi)  \theta (x_\pi - x_t)  \theta (1 - x_\pi).
\end{eqnarray}
and a similar expression for $\bar q_{f,\pi}^t(x_t,Q^2)$. 

Similarly, the contribution of the $\rho$-meson cloud to the structure function is taken into account
 in analogy with the above prescription and the rho structure function is written as~\cite{marco1996}
\begin{equation} \label{F2rho}
F_{2, \rho}^A(x) = - 12 \int d^3 r \int \frac{d^4 p}{(2 \pi)^4}
\theta (p^0) \delta Im D_{\rho} (p) \frac{x}{x_{\rho}} \, 2 M_N
\sum_f e_f^2 x_{\rho}[q_\rho^f(x_\rho(p^0, {\vec p}))+\bar q_\rho^f(x_{\rho}(p^0, {\vec p}))] \theta (1 - x_{\rho})\theta (x_\rho - x)
\end{equation}
and the expression for the rho PDF $q_{f,\rho}^t(x_t,Q^2)$ is derived as~\cite{Marco:1997xb}:
\begin{eqnarray}  \label{F2piqbar}
q_{f,\rho}^t(x_t,Q^2) = - 12 \int  d^3 r  \int  \frac{d^4 p}{(2 \pi)^4} \theta (p^0) \delta I m D_\rho (p)
  2 M_N  q_{f,\rho}(x_\rho)  \theta (x_\rho - x_t)  
\theta (1 - x_\rho),
\end{eqnarray}
\noindent
where $D_{\rho} (p)$ is now the $\rho$-meson propagator in the nuclear medium given by:
\begin{equation}\label{dro}
D_{\rho} (p) = [ {p^0}^2 - \vec{p}\,^{2} - m^2_{\rho} - \Pi^*_{\rho} (p^0, {\bf p}) ]^{- 1}\,,
\end{equation}
where
\begin{equation}\label{pionSelfenergy}
\Pi^*_\rho=\frac{f^2/m_\pi^2 C_\rho F_\rho^2(p)\vec{p}\,^{2}\Pi^*}{1-f^2/m_\pi^2 V'_T\Pi^*}\,.
\end{equation}
Here, $V'_T$ is the transverse part of the spin-isospin interaction, $C_\rho=3.94$, $F_\rho(p)=(\Lambda_\rho^2-m_\rho^2)/(\Lambda_\rho^2+\vec{p}\,^{2})$ is the $\rho NN$ form factor, 
$\Lambda_\rho$=1GeV, $f=1.01$, and $\Pi^*$ is the irreducible rho self energy that contains the contribution of particle - hole and delta - hole excitations and
$\frac{x}{x_{\rho}} = \frac{- p^0 + p^z}{M_N}$. Quark and antiquark PDFs for pions have been taken from the parameterization given by Gluck et al.\cite{Gluck:1991ey} and for the 
rho mesons we have taken the same PDFs as for the pions. It must be pointed out that the choice of $\Lambda_{\pi}$ and $\Lambda_\rho$(=1GeV) in $\pi$NN and $\rho$NN form factors
have been fixed in our earlier works\cite{sajjadnpa,Haider:2015vea,Haider:2016zrk} while describing nuclear medium effects in electromagnetic structure function $F_2^{EM}(x,Q^2)$ to explain the latest data from
JLab and other experiments performed using charged lepton beams on several nuclear targets.

We have also taken into account shadowing effect which arises due to coherent multiple scattering interactions of the intermediate states, which is important in 
DY production at small $x_t$. Various theoretical calculations have indicated that 
 shadowing in DIS as well as in DY processes has a common origin. For the shadowing effect we have followed the model of Kulagin and Petti~\cite{kulagin,Kulagin:2014vsa}.
 
\subsection{Energy loss of beam partons}\label{elossloss}
The incident proton beam traverses the nuclear medium before the beam parton undergoes a hard collision with the target parton. The incident proton may lose energy
due to soft inelastic collisions as it 
 might scatter on its way within the nucleus before producing a lepton pair.

 There are many papers in literature \cite{Marco:1997xb,Johnson2002,Johnson:2000ph,Arleo:2002ph,Garvey2003,Duan2005, Duan:2006hp, Duan:2008qt,Accardi:2009qv,Gavin,vasilev, Kulagin:2014vsa} 
where the effect of energy loss on DY process is discussed and models are given to incorporate them in the calculation of DY yields.  
 However, there is no model which has the preference over the others.
  Most of them perform phenomenological fits and the best value 
  of the parameters are those which have been obtained in the independent analysis of the experimental data. The present situation is summarized by Accardi 
  et al.\cite{Accardi:2009qv}.
  
  For example Duan et al.~\cite{Duan2005, Duan:2006hp, Duan:2008qt} have used two different kinds of quark energy loss expression,
  in which the fractional parton energy $x_b$ is modified to $x_b \rightarrow x_b~+~\Delta x_b$,
   where in the linear fit $\Delta x_b$ are given by
\begin{equation}\label{alpha}
\Delta x_b= {\alpha}\frac{<L>_A}{E_p},
\end{equation}
and by
\begin{equation}\label{aalpha}
\Delta x_b= {\gamma}\frac{<L>^2_A}{E_p}.
\end{equation}
where $<L>_A~\left[=3/4(1.2A^{1/3)}\right]$fm is the average path length
of the incident quark in the nucleus A, $E_p$ is the energy of the
incident proton.
The constants $\alpha$ and $\gamma$ are varied to get a good fit with the experimental data which were found to be in the range of 
 $1.27~\le~\alpha~\le~1.99$ GeV/fm and 
 $0.2~\le~\gamma~\le~0.3$ $GeV/fm^2$~\cite{Duan2005, Duan:2006hp, Duan:2008qt,Johnson2002,Johnson:2000ph,vasilev, Kulagin:2014vsa}. 

 Gavin and Milana~\cite{Gavin} have parameterized the energy loss effect as
\begin{equation}\label{beta}
\Delta x_b= {\beta} x_b A^\frac{1}{3},
\end{equation}
where $\beta=0.0004$.
However, in some recent work of  Johnson et al.~\cite{Johnson2002}, Garvey and Peng~\cite{Garvey2003}, and Kulagin and Petti~\cite{Kulagin:2014vsa}, 
 it has been pointed 
out that a quantitative estimate of energy loss effect in DY processes depends upon how the shadowing effect is treated. 
In the presence of shadowing effect, the fitted parameter for energy loss alpha in equation \ref{alpha} is found to be somewhat smaller in the range of $0.7$ to $1.27$. 
\subsection{Drell-Yan cross sections with nuclear effects}
We have taken into account the various nuclear effects discussed above in this section and write the cross section for the DY process as
 \begin{eqnarray}\label{dycs}
 \frac{d^2 \sigma^{(A)}}{d x_b dx_t} &=&\frac{d^2 \sigma^{(SF)}}{d x_b dx_t}+\frac{d^2 \sigma^{(\pi)}}{d x_b dx_t}+\frac{d^2 \sigma^{(\rho)}}{d x_b dx_t},
\end{eqnarray}
 where $\frac{d^2 \sigma^{(SF)}}{d x_b dx_t}$ is the DY cross section from the nucleons in the nucleus after incorporating
 the nuclear medium effects like Fermi motion, binding energy, nucleon correlations through the use of spectral function. Furthermore, we have also incorporated
 shadowing effect following
 Kulagin and Petti~\cite{Kulagin:2014vsa} and energy loss effect following the phenomenological model given in Eq.~\ref{alpha} with $\alpha$=1.
The expression for  $\frac{d^2 \sigma^{(SF)}}{d x_b dx_t}$ is given by: 
 \begin{eqnarray}\label{spec}
 \frac{d^2 \sigma^{(SF)}}{d x_b dx_t} &=& \frac{4 \pi \alpha^2}{9 q^2} 4 \int d^3 r 
\sum_f e_f^2 \left[q_{f,p} (x_b) \int \frac{d^3 p}{(2 \pi)^3} \frac{M_N}{E (\vec{p})}
\int_{- \infty}^\mu d p^0 S_h (p^0, {\bf p}) \bar{q}_{f,N} (x_t^\prime)\right.\nonumber\\
 &+& \bar{q}_{f,p} (x_b) 
 \left.\int \frac{d^3 p}{(2 \pi)^3} \frac{M_N}{E
(\vec{p})} \int_{ - \infty}^\mu d p^0 S_h (p^0, {\bf p}) 
q_{f,N} (x_t^\prime)\right] \theta (x_t^\prime) \theta (1 - x_t^\prime)~\theta (1 - x_b)
\end{eqnarray}
where $S_h (p^0, {\bf p})$ is the hole spectral function for the nucleon in the nucleus. $q_{f,N}=\frac{1}{2}(q_{f,p}+q_{f,n})$ and
 $\bar q_{f,N}=\frac{1}{2}(\bar q_{f,p}+ \bar q_{f,n})$ are the nucleon PDFs of flavor f averaged over proton
 and neutron in the cases of quarks and antiquarks,respectively.
 
 \begin{figure}
\includegraphics[scale=0.5]{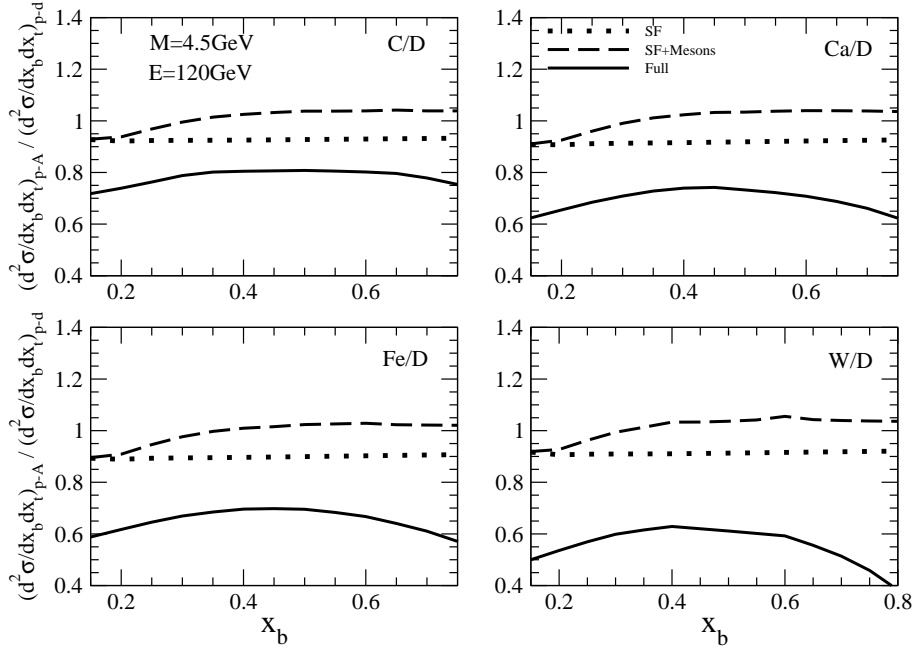}
\caption{$\frac{\left(\frac{d\sigma}{dx_b~dx_t}\right)_{p - ^A}}{\left(\frac{d\sigma}{dx_b~dx_t}\right)_{p - ^{^2}D}}$ vs $x_b$ at $M=4.5GeV$ 
for $A=^{12}C$, $^{40}Ca$, $^{56}Fe$ and $^{184}W$.
 For the beam energy E=120GeV($\sqrt{s_N}$=15GeV) the results are obtained with 
 Spectral function: dotted line, including the mesonic contribution: dashed line, 
 and for the full calculation: solid line. 
 %The results of full calculation at E=800GeV($\sqrt{s_N}$=~38.8GeV) have also been shown: solid line with diamonds.
 }\label{fig13}  
\end{figure}

\begin{figure}
\includegraphics[scale=0.5]{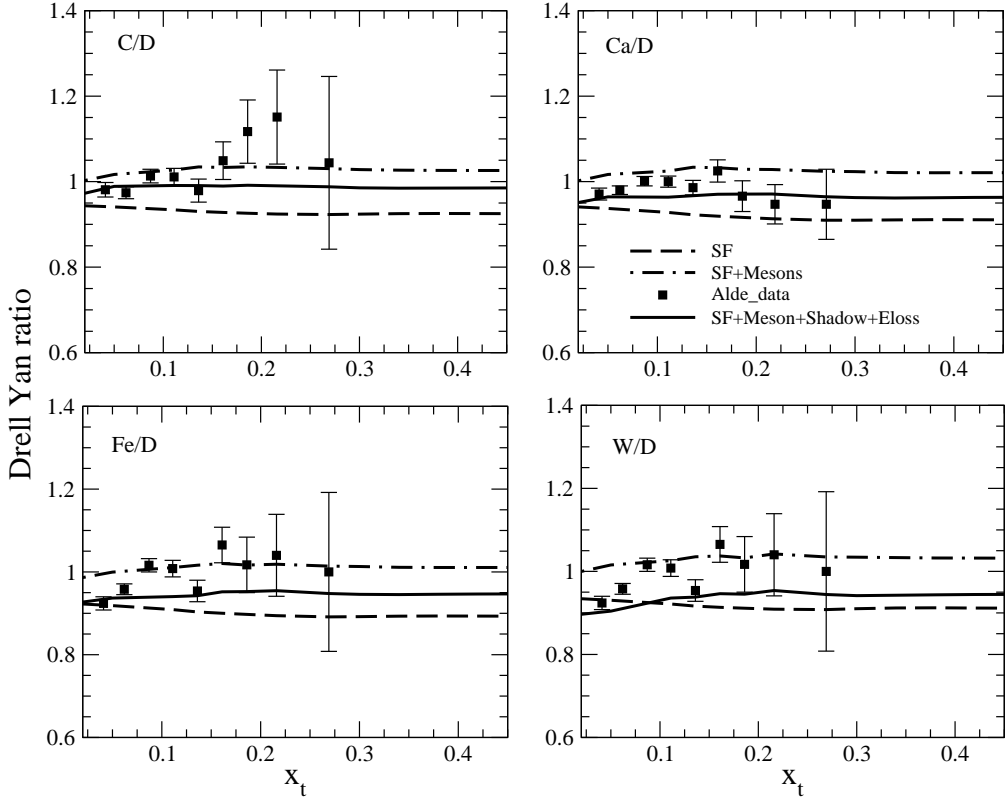}
\caption{Left panel: $\frac{\frac{d\sigma}{dx_t}(C, Fe)}{\frac{d\sigma}{dx_t}(D)}$ vs $x_t$ at E=800GeV($\sqrt{s_N}$=38.8GeV), $x_b=x_t+0.26$, $Q^2 > 16GeV^2$, 
with $\alpha=1$ in Eq.(\ref{alpha}). 
 Spectral function: dashed line, including the mesonic contribution: dashed-dotted line, results of the full calculation: solid line.
 Experimental points are of E772 experiment~\cite{alde}. Right panel: 
$\frac{\frac{d\sigma}{dx_t}(Ca, W)}{\frac{d\sigma}{dx_t}(D)}$ vs $x_t$, lines have same meaning as in the left panel.}
\label{fig7}  
\end{figure}

\begin{figure}
\includegraphics[scale=0.5]{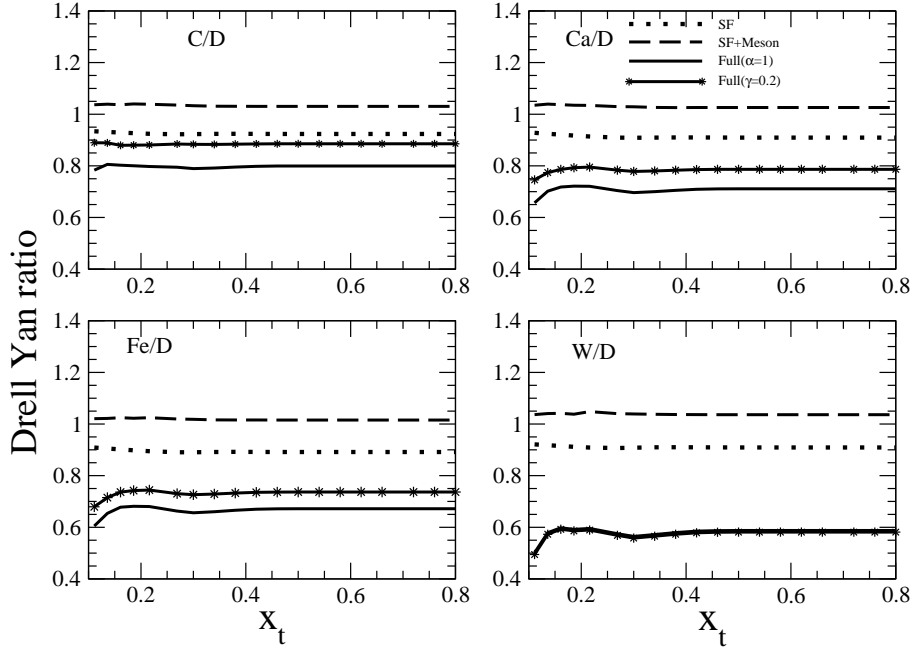}
\caption{Left panel: $\frac{\frac{d\sigma}{dx_t}(C, Fe)}{\frac{d\sigma}{dx_t}(D)}$ vs $x_t$ at  E=120GeV($\sqrt{s_N}$=15GeV), $x_b=x_t+0.26$, $Q^2 > 16GeV^2$.
 The results are obtained with Spectral function: dotted line, 
  Spectral function+Mesonic contribution: dashed line.
  The results of our full calculations are obtained with energy loss using  Eq.(\ref{alpha}) with $\alpha=1$ (solid line) 
  and 
$\gamma=0.2$ in Eq.(\ref{aalpha}) (solid line with stars).
%Experimental points are of E772 experiment~\cite{alde}. 
Right panel: 
$\frac{\frac{d\sigma}{dx_t}(Ca, W)}{\frac{d\sigma}{dx_t}(D)}$ vs $x_t$, lines have same meaning as in the left panel.}
\label{fig77}  
\end{figure}

Similarly to include the pionic contribution $\frac{d^2 \sigma^{(\pi)}}{d x_b dx_t}$ and the rho contribution $\frac{d^2 \sigma^{(\rho)}}{d x_b dx_t}$, 
the DY cross sections are respectively written as~\cite{Marco:1997xb}:
\begin{eqnarray}\label{spec1}
 \frac{d^2 \sigma^{(\pi)}}{d x_b d x_t}
&=& \frac{4 \pi \alpha^2}{9 q^2} (-6) \int d^3 r \sum_f e_f^2 \left[q_{f,p} (x_b) \int \frac{d^4 p}{(2 \pi)^4}
\theta (p^0) \delta Im D (p) 2 M_N \bar{q}_{f,\pi} (x_{\pi})\right.\nonumber\\
 &+& \left.\bar{q}_{f,p} (x_b)\int \frac{d^4 p}{(2 \pi)^4} \theta (p^0)
\delta \, Im \, D (p) 2 M_N q_{f,\pi} (x_{\pi})\right] \theta (x_{\pi} - x_t) \; \theta (1 - x_{ \pi}) \; 
\theta (1 - x_b)
\end{eqnarray}
and 
\begin{eqnarray}\label{specrho}
 \frac{d^2 \sigma^{(\rho)}}{d x_b d x_t}
&=& \frac{4 \pi \alpha^2}{9 q^2} (-12) \int d^3 r \sum_f e_f^2 \left[q_{f,p} (x_b) \int \frac{d^4 p}{(2 \pi)^4}
\theta (p^0) \delta Im D_\rho (p) 2 M_N \bar{q}_{f,\rho} (x_{ \rho})\right.\nonumber\\
 &+& \left.\bar{q}_{f,p} (x_b) \int \frac{d^4 p}{(2 \pi)^4} \theta (p^0)
\delta \, Im \, D_\rho (p) 2 M_N q_{f,\rho} (x_{\rho})\right] \theta (x_{ \rho} - x_t) \; \theta (1 - x_{ \rho}) \; 
\theta (1 - x_b )
\end{eqnarray}
Since in the various experiments the DY cross sections are also obtained in terms of other variables like $M$, $x_f$, $\tau$, etc, where, $M=\sqrt{x_bx_ts_N}$,
$x_f=x_b-x_t$, $\tau=x_bx_t$, therefore, we have also obtained DY cross sections in terms of some of these variables. For example,
using Jacobian transformation Eq.(\ref{spec}) may be written as:
\begin{eqnarray}\label{duan1}
\frac{d^2\sigma}{dx_bdM}&=&\frac{8\pi\alpha^2}{9 M}\frac{1}{x_b s_N} 4 \int d^3 r 
\sum_f e_f^2 \left[q_{f,p} (x_b ) \int \frac{d^3 p}{(2 \pi)^3} \frac{M_N}{E (\vec{p})}
\int_{- \infty}^\mu d p^0 S_h (p^0, {\bf p}) \bar{q}_{f,N} (x_t^\prime)\right.\nonumber\\
&+& \bar{q}_{f,p} (x_b) 
 \left.\int \frac{d^3 p}{(2 \pi)^3} \frac{M_N}{E
(\vec{p})} \int_{ - \infty}^\mu d p^0 S_h (p^0, {\bf p}) 
q_{f,N} (x_t^\prime)\right] \theta (x_t^\prime) \theta (1 - x_t^\prime)~\theta (1 - x_b)
\end{eqnarray}

Most of the experimental results for the DY process have been presented in the form of $\frac{\frac{d\sigma}{dx_bdx_t}(A)}{\frac{d\sigma}{dx_bdx_t}(D)}$ i.e. 
 the ratio of DY cross section in a nuclear target ($\frac{d\sigma}{dx_bdx_t}(A)$) to the DY cross section in deuteron ($\frac{d\sigma}{dx_bdx_t}(D)$). 
 Therefore, to evaluate proton-deuteron DY cross section, we write 
\begin{eqnarray}\label{F2A_Deut1}
\frac{d\sigma^{pd}}{dx_b dx_t} 
&=& 
\frac{d\sigma^{pp}}{dx_b dx_t} + \frac{d\sigma^{pn}}{dx_b dx_t}\,.
\label{eq:dy-pd-exp}
\end{eqnarray}
 and to take into account the deuteron effect, the quark/antiquark distribution function inside the deuteron target have been
calculated using the same formula as for the nuclear structure function but performing the convolution with the deuteron 
wave function squared instead of using the spectral function. The deuteron wave function has been taken from the works of Lacombe et al.~\cite{Lacombe:1981eg}.

In terms of the deuteron wave function, one may write
\begin{eqnarray}\label{F2A_Deut}
q_{f}^t(x_t,Q^2)=\int \frac{d^3p}{(2\pi)^3}\;\frac{M}{E_p^d}\;|\Psi_D({\mathbf p})|^2 {q}_f^{N}(x_t^\prime({\bf p}),Q^2),
\end{eqnarray}
where the four momentum of the proton inside the deuteron 
 is described by $p^\mu=(E_p^d, {\mathbf p})$ with $E_p^d(=M_{\text Deuteron}-\sqrt{M^2+|{\mathbf p}|^2})$ as the energy of the off shell proton 
 inside the deuteron and $M_{\text Deuteron}$ is the deuteron mass. A similar expression has been used for the antiquarks $\bar q_{f}^t(x_t,Q^2)$.
\begin{figure}
\includegraphics[scale=0.5]{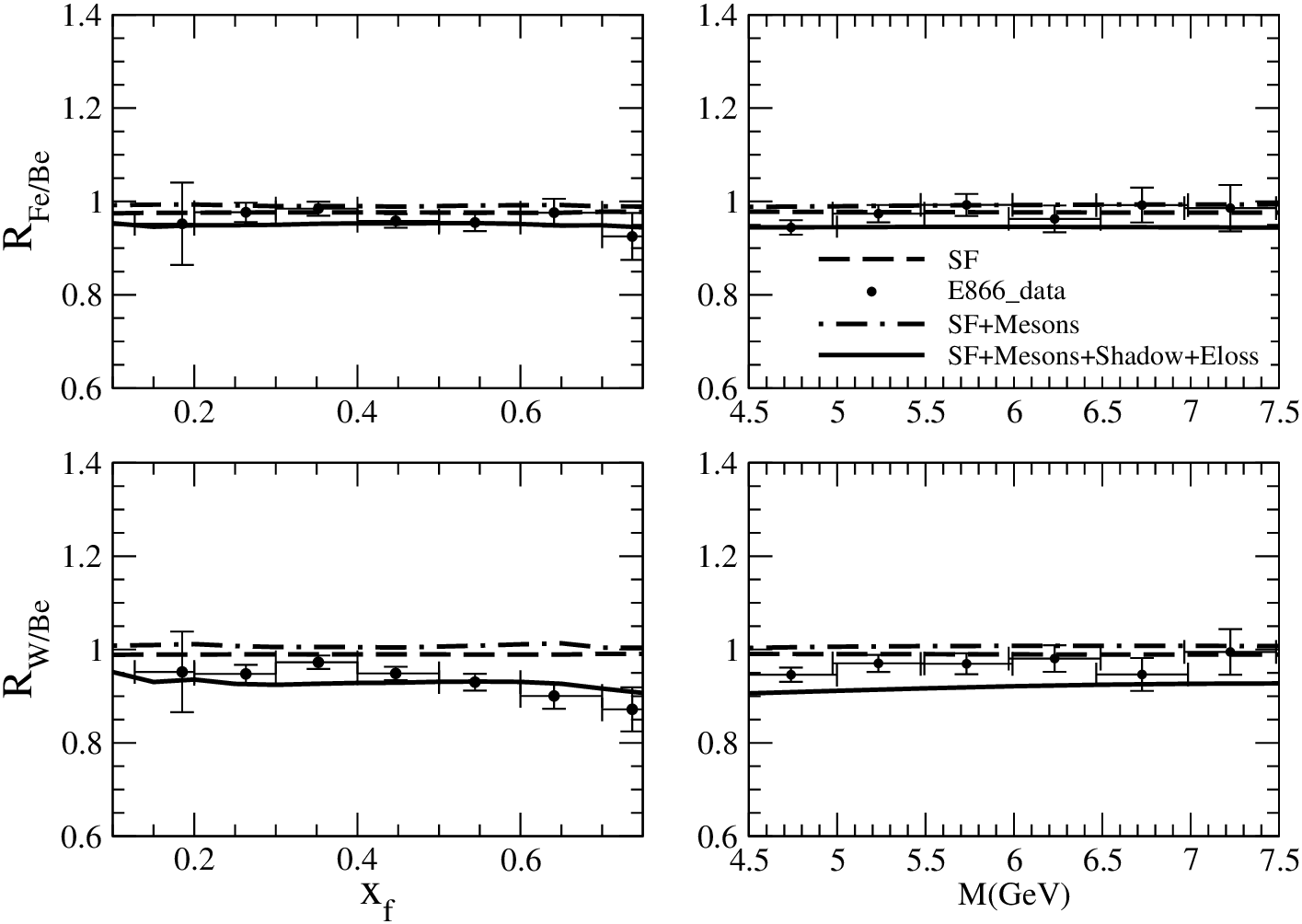}
\caption{ Left Panel: $\frac{\frac{d\sigma}{dx_F}(Fe,W)}{\frac{d\sigma}{dx_F}(Be)}$ vs $x_F$, 
Right Panel: $\frac{\frac{d\sigma}{dM}(Fe,W)}{\frac{d\sigma}{dM}(Be)}$ vs $M(=\sqrt{x_bx_ts_N})$GeV, 
at  E=800GeV($\sqrt{s_N}$=38.8GeV), with $\alpha=1$ in Eq.(\ref{alpha}). Experimental points are of E866 experiment~\cite{e866,vasilev} 
with $0.01 < x_t < 0.12$, $0.21 < x_b < 0.95$ and $0.13 < x_F < 0.93$. 
Spectral function: dashed line, including the mesonic contribution: dashed-dotted line, results of the full calculation: solid line.}\label{fig8}  
\end{figure}

\section{Results and Discussion}\label{sec:RD}

 The results presented here  are based on the following calculations:
 
 (1) DY cross section for proton-nucleus scattering i.e. $\left(\frac{d\sigma}{dx_b dx_t}\right)^A$, where A stands for a nuclear target,
 has been obtained by using the spectral function $S_h (p^0, {\bf p})$ which takes into account Fermi motion,  nucleon correlations and binding energy.  The spectral function 
 with parameters fixed by  Eqs.(\ref{norm2}) and (\ref{norm4}) 
 has been used to calculate the nucleon contribution which reproduce mass number of the nucleon, the binding energy per nucleon for 
 a given nucleus.
 %given in Eq.(\ref{be}), and thus we are left with no free parameter. These have 
 %been discussed in detail in our earlier works~\cite{sajjadnpa,Haider:2015vea}.
 
   (2) We add contributions obtained from the pion cloud using 
    Eq.(\ref{spec1}) and Eq.(\ref{specrho}) for the rho meson to nucleon contributions.
    For evaluating the mesonic contributions the parameters of $D(p)$ in Eq.(\ref{dpi}) and $D_{\rho} (p)$ in Eq.(~\ref{dro}) are 
    fixed by fitting experimental data on $F_2^i(x_t)$ in DIS on various nuclei~\cite{sajjadnpa,Haider:2015vea}.

    (3)  We have also included shadowing effect following the works of Kulagin and Petti~\cite{Kulagin:2014vsa}. With the inclusion of shadowing effect along with the
    spectral function and meson cloud contributions, the numerical results have been presented.

\begin{figure}
\includegraphics[scale=0.5]{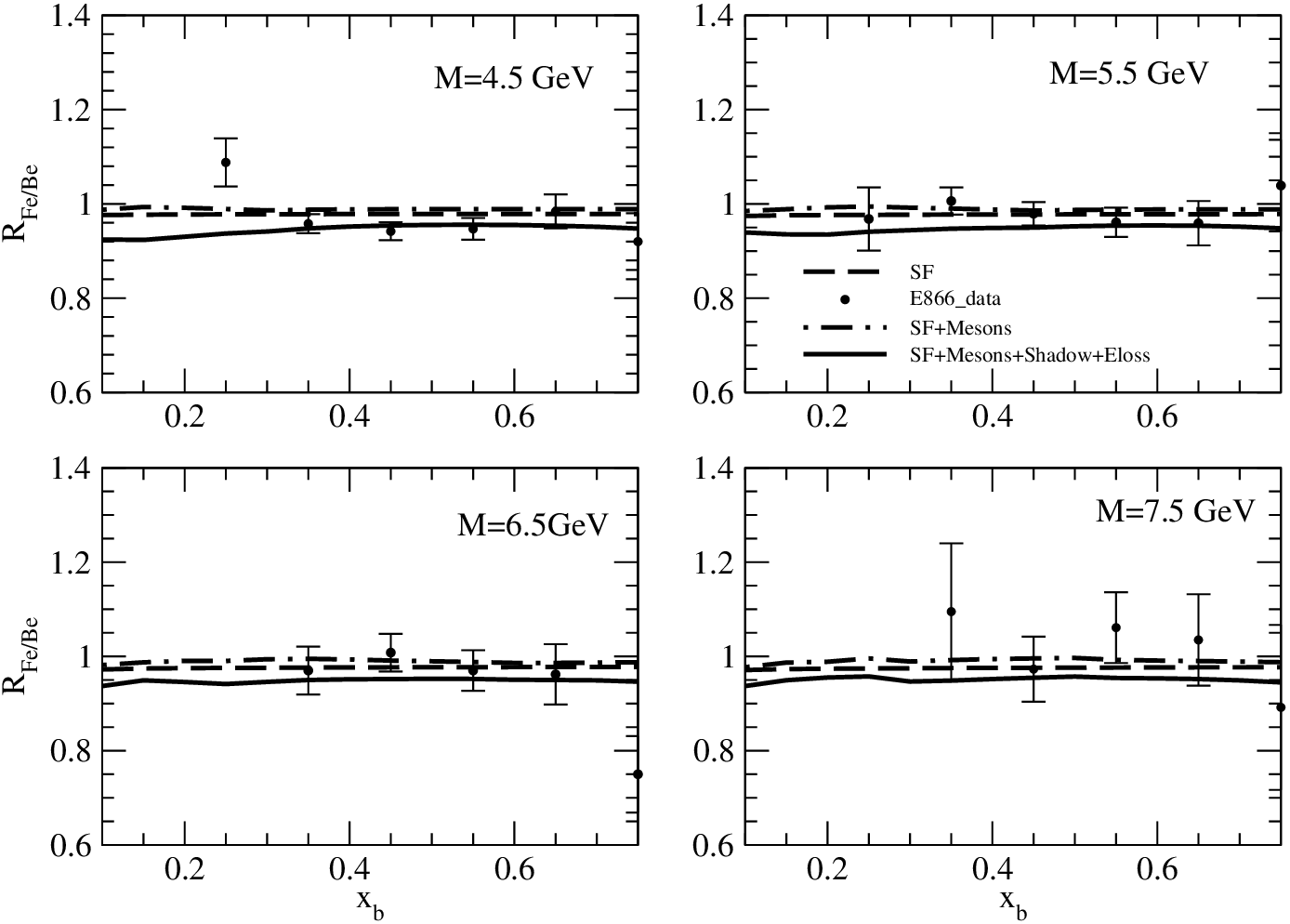}
\caption{$\frac{\frac{d\sigma}{dM dx_b}(Fe)}{\frac{d\sigma}{dM dx_b}(Be)}$ vs $x_b$ at different $M(=\sqrt{x_bx_ts_N})$ with E=800GeV($\sqrt{s_N}$=38.8GeV),
 and $\alpha=1$ in Eq.(\ref{alpha}). 
Experimental points are of E866 experiment~\cite{e866,vasilev}. 
Spectral function: dashed line, including the mesonic contribution: dashed-dotted line, results of the full calculation: solid line}\label{fig9}  
\end{figure}    

    (4) For the energy loss we have used Eq.(\ref{alpha}) with $\alpha=1$. 
      There are other phenomenological models available in the literature to take into account the energy loss effect. 
      We have, therefore, studied the dependence of DY cross sections on energy loss if 
    one uses other phenomenological 
    parameterizatios.
      Some of the expressions are given in 
      Eq.(\ref{aalpha}) and Eq.(\ref{beta}). For our numerical calculations we have taken $\gamma$=0.2. 
    Also we have performed calculations using
    Eq.(\ref{beta}) with $\beta=0.0004$. 
    
   (5) DY cross section for proton-deuteron  scattering has been obtained using Eq.(\ref{F2A_Deut1}) and no energy loss effect has been taken in deuteron.
   We have obtained the results by using Eq.(\ref{F2A_Deut}) with deuteron effect which has found to be small.
 
 (6) For nucleon quark/antiquark PDFs CTEQ6.6~\cite{cteq} has been used and for 
 pion quark/antiquark PDFs parameterization of Gluck et al.~\cite{Gluck:1991ey} has been used.
   
\begin{figure}
\includegraphics[scale=0.5]{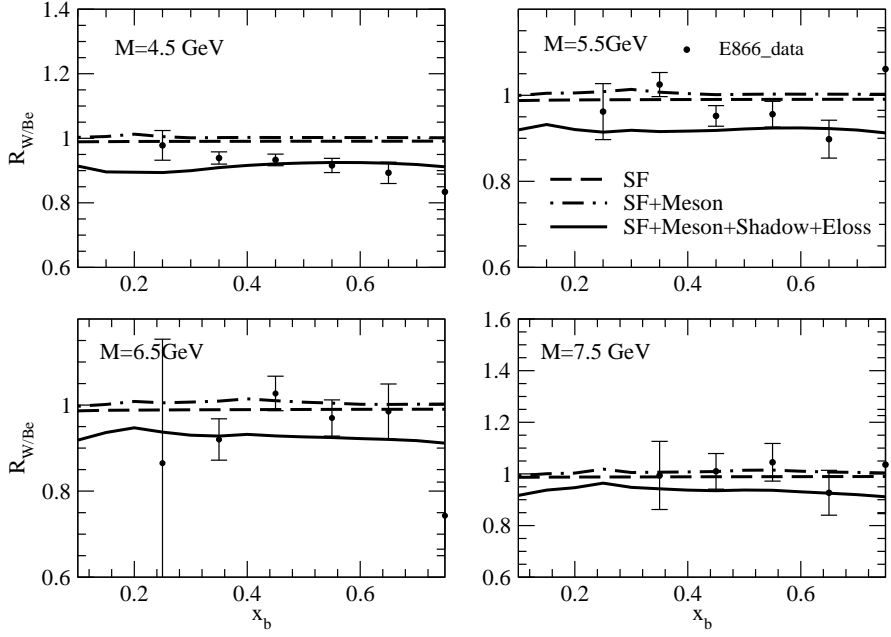}
\caption{$\frac{\frac{d\sigma}{dM dx_b}(W)}{\frac{d\sigma}{dM dx_b}(Be)}$ vs $x_b$ at different $M(=\sqrt{x_bx_ts_N})$ with $\sqrt{s_N}$=38.8GeV, 
with $\alpha=1$ in Eq.(\ref{alpha}). 
Experimental points are of E866 experiment~\cite{e866,vasilev}. 
Spectral function: dashed line, including the mesonic contribution: dashed-dotted line, results of the full calculation: solid line.}\label{fig10}  
\end{figure}

 In Figs.\ref{fig1}-\ref{fig4}, we present the results for the ratio 
 ${\it R}=\frac{\left(\frac{d\sigma}{dx_b~dx_t}\right)_{p - A}}{\left(\frac{d\sigma}{dx_b~dx_t}\right)_{p - ^{^2}D}}$
  vs $x_b$ for M=4.5, 5.5, 6.5 and 7.5GeV. The center of mass energy ($\sqrt{s_N}$) is 38.8GeV. Here A is $^{12}C$ in Fig.\ref{fig1},
  $^{40}Ca$ in Fig.\ref{fig2},  
  $^{56}Fe$ in Fig.\ref{fig3} and 
  $^{184}W$ in Fig.\ref{fig4}. These results are presented for the numerator obtained using the spectral function, including the mesonic effect, and
  also including shadowing and energy loss effect which is our full model. We find that the nuclear structure effects due to bound nucleon lead to a suppression in the DY yield of about  $5-6\%$ in $^{12}C$ in the region 
   of $0.2~<~x_b~<0.6$. 
      This suppression increases with the increase in mass number of the nuclear target. For example, 
  in $^{184}W$ it becomes $6-8\%$ for $0.2~<~x_b~<0.6$.
  Furthermore, we find that there is a significant contribution of mesons which increases the DY ratio i.e. its effect is opposite to the effect 
  of spectral function. For example, the DY yield increases by around $6-8\%$ for $0.2~<~x_b~<0.6$ in $^{12}C$. Moreover, we observe that the effect is more at 
  low $x_b$($\sim 0.2-0.3$) than at high $x_b$. This increase in the DY yield from 
 meson cloud contribution also increases with the mass number A, 
 for example in $^{184}W$ it is around $8-10\%$ for $x_b=0.2-0.3$.  
   We find the contribution from rho meson cloud to be much smaller than the contribution from pion cloud. 
  
  When the shadowing corrections are included there is further suppression in the DY yield and it is effective in the low region of $x_b$.
  The effect of beam energy loss is also to reduce the DY yield. 
  Both effects further adds to the suppression obtained using spectral function, 
  where as mesonic effects lead to an enhancement. The net effect of shadowing and the energy loss effect is  
  $7\%$  at $x_b=0.1$ in $^{12}C$ which becomes $4\%$  at $x_b=0.2$ for $M=4.5GeV$. 
  The shadowing effect as well as energy loss effect are more pronounced in heavier nuclei and suppresses the DY ratio considerably. 
  For example, in $^{184}W$ at low $x_b$ this suppression is about $14-15\%$ at $x_b=0.1-0.2$ which becomes $10\%$ at $x_b=0.3$. Moreover,
  the shadowing effect is $M$ dependent, like in 
  $^{184}W$ the total effect becomes about $10-12\%$ at $x_b=0.1-0.2$ for $M=7.5GeV$.
  It is observed that the suppression in the DY ratio due to energy loss effect (not shown in these 
  figures) is  
   $\sim 2-4\%$ in the case of $^{12}C$ which increases to $3-6\%$ in $^{40}Ca$ and $^{56}Fe$, and becomes  around $4-10\%$ in $^{184}W$ in the region 
   of $x_b~<~0.75$. 
      
    To observe the effect of energy loss using the various approaches, in Fig.\ref{fig66}, we present the results of full calculation
   for $\frac{\left(\frac{d\sigma}{dx_b~dx_t}\right)_{p - A}}{\left(\frac{d\sigma}{dx_b~dx_t}\right)_{p - ^{^2}D}}$ at M=4.5GeV
   and $\sqrt{s_N}$=38.8GeV for
 $\alpha$=1 in Eq.~(\ref{alpha}), $\gamma$=0.2 in Eq.~(\ref{aalpha}) and $\beta=0.0004$ in Eq.(\ref{beta}). We find that 
 there is hardly any difference in the results obtained with $\gamma$=0.2 in Eq.~(\ref{aalpha}) in comparison 
 to the results obtained with $\alpha$=1 in Eq.~(\ref{alpha}). When we obtain the results using $\beta=0.0004$ in Eq.(\ref{beta}), 
 it is found that there is less reduction in the DY ratio for heavier nuclear targets like $^{184}W$ as compared to 
 the results obtained using  $\alpha$=1 in Eq.~(\ref{alpha}), while for the light nuclear targets the results are comparable.  We have also 
 studied(not shown here) the dependence of the parameter $\alpha$ used in the expression given in Eq.(\ref{alpha}) for energy loss, we have
 varied $\alpha$ in the range
    $0.5~<~\alpha~<~3$, corresponding to the range of values used in literature~\cite{Duan2005,Duan:2006hp,Johnson2002},
    for $\frac{\left(\frac{d\sigma}{dx_b~dx_t}\right)_{p - A}}{\left(\frac{d\sigma}{dx_b~dx_t}\right)_{p - ^{^2}D}}$ 
    at M=4.5GeV and $\sqrt{s_N}$=38.8GeV. We find that in the case of p-$^{12}C$ DY process, there is less(more) 
 suppression when $\alpha$ is taken as 0.5(2) from the results obtained at $\alpha$=1, our reference value. For example, this 
 is around $1-2\%$ lesser(larger) at $x_b \sim$ 0.1-0.2 at $\alpha$=0.5(2).
 With the increase in mass number the difference in the results increases. For example,  in $^{184}W$ 
 this suppression is around $2-3\%$ lesser(larger) at $x_b \sim$ 0.1-0.2 at $\alpha$=0.5(2).
 Thus we observe that with the increase in $\alpha$, DY ratio decreases considerably and there is an 'A' dependence on the DY ratio. 
 The numerical results are compared with the E772 experimental data and found to be in fair agreement.
 
 To explicitly compare the effect of nuclear medium as well as energy loss in DY production cross section at different center of mass energies,
        in Fig.~\ref{fig13}, we have presented the results for 
      $\frac{\frac{d\sigma}{dx_b dx_t}(A)}{\frac{d\sigma}{dx_b dx_t}(D)}$ vs $x_b$ at $M=4.5GeV$ for $A=^{12}C$, $^{40}Ca$, $^{56}Fe$ and $^{184}W$ 
   at E=120GeV($\sqrt{s_N}$=15GeV).
   These results depict how the DY ratio vary
   at the different center 
 of mass energies when the results are obtained with Spectral function, including mesonic contribution and the full calculation without energy loss
  and with energy loss for $\alpha=1$ in Eq.(\ref{alpha}).  For E=120GeV,
  we observe that the effect of spectral function is to reduce DY yield by about $7\%$ for $^{12}C$ and the reduction increases with the increase in mass number like 
  for $^{184}W$ it is $9\%$. When the meson cloud contributions are included the DY yield get enhanced by $4-6\%$ at low $x_b$ and $10-12\%$ at 
  mid and high values of $x_b$ and is found almost independent of $A$. When shadowing and energy loss effects are added, the results get reduced by $13-15\%$
  in the range of $0.2<x_b<0.6$ and show significant reduction at high values of $x_b$ which is around $16-28\%$  in $^{12}C$.
 Furthermore, we observe a strong nuclear mass dependence on the DY yield due to energy loss and shadowing effects, for example it is $45-50\%$
  for $^{184}W$. 
  When we compare the present results obtained using full model with the results obtained at E=800GeV for M=4.5GeV(Figs.\ref{fig1}-\ref{fig4}),
  it may be observed that the decrease in DY yield is mainly due to the large energy loss at low center of mass energy. 
 
 In Fig.\ref{fig7}, we present the results 
   for $\frac{\left(\frac{d\sigma}{dx_t}\right)_{p - A}}{\left(\frac{d\sigma}{dx_t}\right)_{p - ^{^2}D}}$ vs $x_t$ at $\sqrt{s_N}$=38.8GeV
   by integrating over $x_b$. The integration over $x_b$ is done by putting the constraints as 
   $x_F(=x_b-x_t)~> ~0.26$ and $Q^2 > 16GeV^2$.  We find the effect of spectral function to be around $6-8\%$ in $^{12}C$ 
   which increases with mass number and becomes around $8-9\%$ in $^{184}W$, and is found to be almost independent of $x_t$.
   The addition of the mesonic contribution enhances the DY yield, for example, 
   in $^{12}C$ this increase is $10-12\%$ which further get enhances with increase in mass number and becomes around $12-14\%$ in $^{184}W$.
   When the results are obtained with shadowing and the beam energy loss effect the suppression is the numerical results in $^{12}C$ is around $3-4\%$, and it increases with mass number and 
   becomes around $8-9\%$ in $^{184}W$. The present theoretical results are also compared with E772~\cite{alde} experimental data. 
    
     In view of the E906 SeaQuest experiment being done at Fermi Lab, we have presented the results in Fig.~\ref{fig77}, for $\frac{\left(\frac{d\sigma}{dx_t}\right)_{p - A}}{\left(\frac{d\sigma}{dx_t}\right)_{p - ^{^2}D}}$ vs $x_t$, 
     ($A$=$^{12}C$,~$^{40}Ca$,~$^{56}Fe$~and~$^{184}W$), at $\sqrt{s_N}$=15GeV corresponding 
     to the energy of the incident proton E=120GeV. We find that the effect of spectral function to be around $7-8\%$ in $^{12}C$ 
   which increases with mass number and becomes around $8-9\%$ in $^{184}W$. When mesonic effects are included the rise in the DY ratio is around 
   $11-12\%$ in $^{12}C$ 
   which increases with mass number and becomes around $13-14\%$ in $^{184}W$. When shadowing and energy loss effects are further added, then there is 
   large reduction which is mainly due to loss of beam energy at low $\sqrt{s}$, which increases considerably with the increase in mass number of target nuclei.
   For example, for $\alpha$=1 in Eq.~(\ref{alpha}) the results  are reduced by  $22-24\%$ in $^{12}C$, $30-32\%$ in $^{40}Ca$ and $^{56}Fe$, 
     and $42-44\%$ in $^{184}W$. To see the dependence of the different approaches of energy loss effect on the DY ratio, we have also obtained 
     the results using $\gamma=0.2$ in Eq.~(\ref{aalpha}), and find that for low mass nuclei there is difference in the results of 
     about 8-10$\%$ in $^{12}C$, which becomes negligible 
     with the increase in mass number of nuclear target. 
     
 In E866 experiment~\cite{e866,vasilev}, the results were obtained for $\frac{d\sigma}{dx_F}$ vs $x_F$, where $x_F=x_b - x_t$ 
 and  $\frac{d\sigma}{dM}$ vs $M$, with $M=\sqrt{x_bx_ts_N}$. 
 Using Eqs.\ref{spec} and \ref{duan1}, we have obtained the results respectively for $\frac{d\sigma}{dx_F}$ vs $x_F$ and 
 $\frac{d\sigma}{dM}$ vs $M$ and shown these results in 
 Fig.\ref{fig8}. For $\frac{d\sigma}{dx_F}$ vs $x_F$, 
 we have integrated over $x_b$ between the limits $0.21 \le x_b \le 0.95$ and followed the kinematical cuts of $4.0<M<8.4$ GeV used in the analysis of
 E866~\cite{e866,vasilev} 
 experiment. In the case of $\frac{d\sigma}{dM}$ vs $M$, we have integrated over $x_b$ between the limits $0.21 \le x_b \le 0.95$ and put the
 kinematical constraint $0.13 \le x_F \le 0.93$ as used in E866~\cite{e866,vasilev} experiment. 
 These results are shown for the DY ratio for $\frac{\left(\frac{d\sigma}{dx_F}\right)^{i}}{\left(\frac{d\sigma}{dx_F}\right)^{Be}}$ vs $x_F$(Left panel) and 
$\frac{\left(\frac{d\sigma}{dM}\right)^{i}}{\left(\frac{d\sigma}{dM}\right)^{Be}}$ vs $M$(Right panel), where i stands for iron(top panel) and tungsten(bottom panel)
 nuclei. These results are obtained with spectral function, including mesonic effect, and using our full model with energy loss effect. For the energy loss, we have used Eq.(\ref{alpha}) with $\alpha=1$. 
 We observe that the effect of spectral function($\sim~1\%$) and meson cloud contributions($\sim~2\%$) are small. When shadowing and energy loss 
 effects are included there is a significant reduction in DY yield which is around $4-5\%$ in $\frac{Fe}{Be}$ and the reduction increases to
 $8-10\%$ in $\frac{W}{Be}$. 
 We find a good agreement with the experimental results for the various DY ratios available from E866~\cite{e866,vasilev} experiment. 

 In Fig. \ref{fig9}, we present the results of DY ratio for 
 $\frac{\left(\frac{d^2\sigma}{dx_bdM}\right)^{Fe}}{\left(\frac{d^2\sigma}{dx_bdM}\right)^{Be}}$
 vs $x_b$ for different values
 of $M(=\sqrt{x_bx_ts_N})$, between the kinematic limits $0.13 \le x_F \le 0.93$ and 
 $0.21 \le x_b \le 0.95$ as used in E866 experiment~\cite{dyhepdata,web}. 
 The results of this ratio for $\frac{\left(\frac{d^2\sigma}{dx_bdM}\right)^{W}}{\left(\frac{d^2\sigma}{dx_bdM}\right)^{Be}}$ are shown in  Fig. \ref{fig10}.
 The results are presented to observe the effect for spectral function, meson cloud contributions, and with shadowing effect and energy loss effect on the DY production.

\section{Summary and Conclusion}\label{sec:CC}
We have studied nuclear medium effects in DY process using quark parton distribution functions and nucleon structure functions for a bound nucleon. 
 We have used a microscopic nuclear model which takes into account the effect of Fermi
motion, nuclear binding and nucleon correlations through a relativistic spectral function of bound nucleon. 
 The contributions of $\pi$ and $\rho$ mesons are also included. Furthermore, shadowing corrections are taken into account.
We have also included the beam energy loss effect due to initial state interactions of protons with nuclear 
constituents before they suffer 
hard collisions to produce lepton pair. 
 We find a reduction in the DY yield due to nuclear structure effects and an enhancement due to mesonic contribution. 
 The effect of shadowing is to reduce the DY yield in the 
 region of very low $x_t$($x_t~<~0.15$). Both the reduction as well as the enhancement
 in the case of DY yields are found to be of similar magnitude as in the case of DIS of charged leptons. 
 In the case of DY yields there is a further reduction due to beam energy loss effect in the nuclear medium
 which has been treated phenomenologically using a parameter describing the beam energy loss. The numerical results are compared with 
 the experimental results from E772~\cite{alde} and E866~\cite{e866,vasilev} experiments.
 A reasonable agreement with the experimental results presently available for $^{12}C$, $^{40}Ca$, $^{56}Fe$, and $^{184}W$ 
 has been found. We have also presented in this paper, results for $\frac{d^{2}\sigma}{dx_bdx_t}$ vs $x_b$ for various
 values of $M(=\sqrt{x_bx_ts_N})$ and the results for  $\frac{d\sigma}{dx_t}$ vs $x_t$  relevant to the 
forthcoming E906 SeaQuest~\cite{Seaquest} experiment at Fermi Lab. 
Our results show that the model for describing the nuclear medium effects in the DIS of charged leptons and 
neutrino and antineutrino with nuclear targets is able to explain the experimental results in the case of DY yield in the region 
$0.1 < x_t < 0.35$. High statistics, high precision data from E906 SeaQuest~\cite{Seaquest} 
experiment on $\frac{d^{2}\sigma}{dx_bdx_t}$ in various regions of $x_b$ and $x_t$ will provide 
important information about the modification of quark PDFs and nucleon structure function in the nuclear medium.
\section{Acknowledgments}
M. S. A. is thankful to Department of Science
and Technology(DST), Government of India for providing financial assistance under Grant No. SR/S2/HEP-18/2012. I.R.S. thanks FIS2014-59386-P Spanish project for financial support and Juan de la Cierva-incorporacion contract from Spanish MINECO.

\end{document}